# Structural Semiconductor-to-Semimetal Phase Transition in Two-Dimensional Materials Induced by Electrostatic Gating

Yao Li[1], Karel-Alexander N. Duerloo[2], Kerry Wauson[3], Evan J. Reed[2][1]

[1]Department of Applied Physics, Stanford University, Stanford, CA 94305, United States

[2]Department of Material Science and Engineering, Stanford University, Stanford, CA 94305, United States

[3]Klipsch School of Electrical and Computer Engineering, New Mexico State University, Las Cruces, NM 88003, United States

Dynamic control of conductivity and optical properties via atomic structure changes is of tremendous technological importance in information storage. Energy consumption considerations provide a driving force toward employing thin materials in devices. Monolayer transition metal dichalcogenides are nearly atomically-thin materials that can exist in multiple crystal structures, each with distinct electrical properties. Using density functional approaches, we discover that electrostatic gating device configurations have the potential to drive structural semiconductor-to-semimetal phase transitions in some monolayer transition metal dichalcogenides. For the first time, we show that the dynamical control of this phase transition can be achieved in carefully designed electronic devices. We discover that the semiconductor-to-semimetal phase transition in monolayer $MoTe_2$ can be driven by a gate voltage of several Volts with appropriate choice of dielectric. Structural transitions in monolayer $TaSe_2$ are predicted to occur under similar conditions. While the required field magnitudes are large for these two materials, we find that the gate voltage for the transition can be reduced arbitrarily by alloying, e.g. for $Mo_xW_{1-x}Te_2$ monolayers. We have developed a method for computing phase diagrams of monolayer materials with respect to charge and voltage, validated by comparing to direct calculations and experimental measurements. Our findings identify a new physical mechanism, not existing in bulk materials, to dynamically control structural phase transitions in two-dimensional materials, enabling potential applications in phase-change electronic devices.

---
[1] To whom correspondence should be addressed:
Email: evanreed@stanford.edu
Tel: [+1] (650) 723-2971
Fax: [+1] (650) 725-4034
URL: http://www. stanford.edu/group/evanreed
Postal: 496 Lomita Mall, Stanford, CA 94305, USA

**Background**

Structural phase transitions yielding a change of electrical conductivity are a topic of long-standing interest and importance[1,2]. Two of the most studied phase-change material classes for electronic and optical applications are metal-oxide materials[3,4] and GeSbTe (GST) alloys[5], both having a large electrical contrast. For example, the metal-oxide material vanadium dioxide ($VO_2$) is reported to exhibit a structural metal-insulator transition near room temperature at ultrafast timescales, which can be triggered by various stimuli including heating[6], and optical[7] excitations, and strain[8]. GST alloys can undergo reversible switching between amorphous and crystalline states with different electrical resistivity and optical properties. This is usually achieved by Joule heating employed in phase change memory applications[9,10]. These materials are distinguished from the myriad materials that exhibit atomic structural changes by the proximity of a phase boundary to ambient conditions.

Another group of materials that can undergo phase transitions are layered transition metal dichalcogenides (TMDs) which have received recent attention as single and few-layer materials, although research on bulk TMDs dates back decades[11,12]. Early attention has been focused primarily on electronic transitions between incommensurate and commensurate charge density wave (CDW)[13,14] phases and superconducting phases[15]. Some TMDs have been found to exist in multiple crystal structures[16], and transitions between them have been demonstrated in group V TMDs ($TaSe_2$ and $TaS_2$) utilizing an STM tip[17,18]. These reported transitions in $TaSe_2$ and $TaS_2$ are between two metallic phases. Recently group VI TMDs have attracted increasing attention because they can exist in a semiconducting phase[19]. Recent computational work indicates that structural transitions between phases of large electrical contrast in some exfoliated two-dimensional (2D) group VI TMDs can be driven by mechanical strain[20]. Excess charges transferred from chemical surroundings are also reported to induce structural phase transitions in 2D group VI TMDs[21–24]. One would like to know the threshold charge density required to induce these transitions and whether these transitions could be dynamically controlled by electrostatic gating, utilizing standard electronic devices.

In this work we use density functional theory (DFT) to determine the phase boundaries of single layer $MoS_2$, $MoTe_2$, $TaSe_2$, and the alloy $Mo_xW_{1-x}Te_2$. We consider $MoS_2$ because it has received considerable attention as an exceptionally stable semiconductor, and $MoTe_2$ because DFT calculations indicate that the energy difference between semiconducting and semimetallic phases is exceptionally small among Mo- and W- TMDs[20]. We calculate the phase boundaries at conditions of constant charge and constant voltage, the electrical analogs to mechanical conditions of constant volume and constant pressure, respectively. We find that a surface charge density of less than -0.04 $e$ or greater than 0.08 $e$ per formula unit is required to observe the semiconductor-to-semimetal phase transition in monolayer $MoTe_2$ under constant stress conditions ($e$ is the elementary electric charge) and a much larger value of approximately -0.33 $e$ or 0.92 $e$ per formula unit is required in the monolayer $MoS_2$ case. The charge densities discussed in this work refer to excess charge density and should not be misinterpreted as the electron or hole density in a charge-neutral material that one might obtain from chemical doping.

We also study the potential of phase control in monolayer $MoTe_2$ through electrostatic gating using a capacitor structure. We discover that a gate voltage as small as a few Volts for some choices of gate dielectric can be applied to drive the phase transition in monolayer $MoTe_2$ using a capacitor structure. While the required field magnitudes are large and may be challenging to achieve, we find that the transition gate voltage may be reduced to 0.3–1 V and potentially lower by substituting a specific fraction of W atoms within MoTe$_2$ monolayers to yield the alloy Mo$_x$W$_{1-x}$Te$_2$. To accomplish these calculations, we have developed a DFT-based model of the electrostatically gated structure. This approach is validated by comparing to direct DFT calculations and by comparing the computed phase boundary of the structural phase transition in TaSe$_2$ to earlier reported STM experimental results[17] where reasonable agreement is found.

**Crystal structures**

TMDs are a class of layered materials with the formula MX$_2$, where M is a transition metal atom and X is a chalcogen atom. Each monolayer is composed of a metal layer sandwiched between two chalcogenide layers, forming a X-M-X structure[16] that is three atoms thick. The weak interlayer attraction of TMDs allows exfoliation of these stable three-atom-thick layers. Given the crystal structures reported in the bulk, we expect that exfoliated monolayer TMDs have the potential to exist in the crystal structures shown in Figure 1. Figure 1 shows the X atoms with trigonal prismatic coordination, octahedral coordination, or a distorted octahedral coordination around the M atoms[16,20,25]. We will refer to these three structures of the monolayer as the 2H phase, 1T phase, and 1T' phase, respectively. Symmetry breaking in the 1T' leads to a rectangular primitive unit cell.

Among these 2D TMDs, the Mo- and W-based materials have attracted the most attention because their 2H crystal structures are semiconductors with photon absorption gaps in the 1-2 eV[26] range, showing potential for applications in ultrathin flexible and nearly transparent 2D electronics. Radisavljevic et al. fabricated single-layer MoS$_2$ transistors of high mobility, large current on/off ratios and low standby power dissipation[27]. Unlike group IV and group V TMDs (for example TaSe$_2$, TaS$_2$) which have been observed in the metallic 1T crystal structure[16], DFT calculations on the group VI TMD (Mo- and W-based) freestanding monolayers indicate that the 1T structure is unstable in the absence of external stabilizing influences[20]. However, group VI TMDs do have a stable octahedrally-coordinated structure of large electrical conductivity, which is a distorted version of the 1T phase and referred to as 1T' structure (Figure 1). Based on DFT calculation results, Kohn-Sham states of this 1T' crystal structure have metallic or semimetallic characteristics, consistent with previous experiments[16]. This octahedral-like 1T' crystal structure has been observed in WTe$_2$ under ambient conditions[16,28], in MoTe$_2$ at high temperature[28], and in Lithium-intercalated MoS$_2$[25]. There is recent experimental evidence that few layer films of the T' phase of MoTe$_2$ exhibit a bandgap that varies from 60 meV to zero with variations in number of layers[29].

The relative energies of Mo- and W-based TMD monolayer crystals shown in Figure 1 have been calculated using semilocal DFT with spin-orbit coupling, as shown in Supplementary Figure 5. These results are consistent with experimental evidence that the bulk form of WTe$_2$ is stable in the metallic 1T' phase, while other Mo- and W-dichalcogenides are stable in the semiconducting 2H phase[16]. These calculations indicate that the switch from semiconducting 2H phase to semimetallic 1T' phase in monolayer MoTe$_2$ requires the least energy (31 meV per formula unit), suggesting the potential for a transition that is exceptionally close to ambient conditions. Therefore we choose to focus on determining the phase boundary of monolayer MoTe$_2$. While the computed energy difference between 2H and 1T' is considerably larger for MoS$_2$ (548 meV per formula unit), we also compute phase boundaries for this monolayer at constant charge because it has received more attention in the laboratory to date. Among 2D group VI TMDs, monolayer MoS$_2$ has attracted the most experimental attention for its stability and relative ease of exfoliation and synthesis. Monolayer MoTe$_2$ has also been exfoliated[30,31] and its synthesis is a fast-developing field.

**Energy calculations for systems containing a charged monolayer**

We examine two distinct thermodynamic constraints for a system containing a charged monolayer. In one scenario, the monolayer is constrained to be at a constant excess charge per unit area, as shown in Figure 2a; in the other, the monolayer is constrained to be at constant voltage, as shown in Figure 2b. These are the electrical analogs to mechanical conditions of constant volume and constant pressure, respectively. Layer I is a monolayer TMD with a Fermi level $\mu_f^I$, and plate II has a Fermi level of $\mu_f^{II}$. A dielectric medium of thickness $d$ and capacitance $C$ is sandwiched between monolayer TMDs and plate II. This dielectric medium can be vacuum. Distance $s^I$ is the separation between the center of monolayer TMD I and the left surface of the dielectric medium, while $s^{II}$ is the separation between the surface atoms of plate II and the right surface of the dielectric medium. (See Supplementary Information for more details about distance parameters).

When the charge Q on the monolayer is fixed, the total energy required to move charge Q from plate II to the monolayer TMD $E(Q)$ is the sum of three parts: energy stored in the dielectric medium ($E_c$), energy of moving electrons Q from the Fermi level of plate II to the dielectric surface ($E^{II}$), and energy of moving electrons Q from the other dielectric surface to the Fermi level of monolayer TMDs ($E^{I}$), as shown in Figure 2.

$$E(Q) = E^I(Q, s^I) + E^{II}(-Q, s^{II}) + E_c = E^I(Q, s^I) + E^{II}(-Q, s^{II}) + \frac{Q^2}{2C} \qquad (1)$$

where $C$ is the capacitance of the dielectric medium.

The first term in equation (1), $E^I(Q, s^I)$, is calculated using DFT as described in Supplementary Information section 1. We take plate II to be a bulk metal with a work function $W$ so that the second term in equation (1) can be approximately written as:

$$E^{II}(-Q, s^{II}) = -QW \qquad (2)$$

When the voltages is fixed rather than the charge, the grand potential $\Phi_G(Q, V)$ becomes the relevant thermodynamic energy defined as:

$$\Phi_G(Q, V) = E(Q) - QV \qquad (3)$$

where $E(Q)$ is computed using equation (1). The $QV$ term in this expression represents external energy supplied to the system when the charge $Q$ flows through an externally applied voltage $V$. The equilibrium charge $Q_{eq}$ can be calculated through minimization of the grand potential at a given gate voltage $V$.

$$\left.\frac{\partial \Phi_G(Q,V)}{\partial Q}\right|_{Q=Q_{eq}} = 0 \qquad (4)$$

Applying computed $Q_{eq}(V)$ to equation (3), we can obtain the equilibrium grand potential as a function of gate voltage $\Phi_G^{eq}(V)$.

$$\Phi_G^{eq}(V) = \Phi_G\big(Q_{eq}(V), V\big) = E\big(Q_{eq}(V)\big) - Q_{eq}(V)V \qquad (5)$$

Hereafter, we omit the superscript "eq" for the equilibrium grand potential $\Phi_G^{eq}(V)$.

In addition to the electrical constraint, the nature of the mechanical constraint on the monolayer is also expected to play a role in the phase boundary, discussed in section 2 of the supplementary information.

**Phase boundary at constant charge**

The distinction between the constant charge and voltage cases is most important when a phase transformation occurs. We discover that the transition between semiconducting 2H-TMDs and semimetallic 1T'-TMDs can be driven by excess electric charge (positive or negative) in the monolayer. A constant charge condition exists when the charge on the monolayer remains constant during the phase transition as if it is electrically isolated. An approximate condition of constant charge could exist when adsorbed atoms or molecules donate charge to the monolayer.

Figure 3 presents the energy difference between the 2H and 1T' phases as a function of the charge density in the monolayer. Figure 3a is a schematic of the system showing relevant distances used in our computations. The monolayer TMD is a distance $d$ away from the electron reservoir (metal electrode). More information on the choice of distance parameters is given in Supplementary Information. The energy of such a system is described by equation (1). When computing the energy difference between a system where the monolayer is in the 2H phase and another system where the monolayer is in the 1T' phase $E_H(Q) - E_{T'}(Q)$, the terms $E^{II}(-Q, s^{II})$ and capacitance terms $E_c$ in equation (1) cancel, leading to,

$$E_H(Q) - E_{T'}(Q) = [E^I(Q, s^I) + E^{II}(-Q, s^{II}) + E_c]_H - [E^I(Q, s^I) + E^{II}(-Q, s^{II}) + E_c]_{T'} = E^I{}_H(Q, s^I) - E^I{}_{T'}(Q, s^I) \qquad (6)$$

The blue lines are constant-stress cases, in which both phases exhibit minimum energy lattice constants and atomic positions. This condition is expected to hold when the monolayer is freely suspended or is not constrained by friction on a substrate. The red lines represent constant-area cases, where the monolayer is clamped to its 2H lattice constants. This condition might be expected to hold when there is a strong frictional interaction between the monolayer and substrate preventing the monolayer from relaxing freely.

Figure 3b shows that semiconducting 2H-MoTe$_2$ has lower free energy and is the equilibrium state when the monolayer is electrically neutral or minimally charged. For the stress-free case (blue line), when the charge density is between -0.04 $e$ and 0.08 $e$ per formula unit, 2H-MoTe$_2$ is the thermodynamically stable phase. These charge densities correspond to $-1.8 \times 10^{13}$ and $3.6 \times 10^{13}$ $e$/cm$^2$, respectively. Outside this range, semimetallic 1T'-MoTe$_2$ will become the equilibrium phase and a transition from the semiconducting 2H phase to the semimetallic 1T' phase will occur.

In the constant-area case (red line) in Figure 3b, a considerably larger charge density is required to drive the phase transition. This suggests that the precise transition point may be sensitive to the presence of a substrate and that the detailed nature of the mechanical constraint of the monolayer may play a substantive role in the magnitude of the phase boundaries. The higher transition charge in this case can be understood by considering that the energy of the strained T' phase is higher than that of the zero stress T' phase, pushing the phase boundary to larger charge states.

Figure 3c shows that the transition in monolayer MoS$_2$ requires much larger charge density than the MoTe$_2$ case. If the negative charge density is larger than 0.33 $e$ per MoS$_2$ formula unit, semimetallic 1T'-MoS$_2$ will have lower free energy and be more stable. For negative charge densities less than 0.33 $e$ per formula unit, semiconducting 2H-MoS$_2$ will be energetically favorable. This is consistent with previous experimental reports that adsorbed species donating negative charge to monolayer MoS$_2$ can trigger a trigonal prismatic to octahedral structure transformation[24,32]. MoS$_2$ single layers are reported to adopt a distorted octahedral structure when bulk MoS$_2$ is first intercalated with lithium to form Li$_x$MoS$_2$ with $x \approx 1.0$ and then exfoliated by immersion in distilled water.[25] This is consistent with our prediction that a negative charge density larger than 0.33 *e* per MoS$_2$ may trigger the phase transition from 2H phase to 1T' phase MoS$_2$. If the charging is positive, a substantially larger charge density is needed to drive the transition.

**Phase boundary at constant voltage**

Another relevant type of electrical constraint is fixed voltage or electron chemical potential. This constraint is most applicable when the monolayer is in an electrostatic gating structure similar to field-effect transistors made using monolayers. Such a device structure enables a dynamical approach to achieve semiconductor/semimetal phase control in monolayer TMDs, suggesting intriguing applications for ultrathin flexible 2D electronic devices including phase change memory.

Many distinct electrostatic gating device structures can be utilized to realize this dynamic control through a change in carrier density or electron chemical potential of the monolayer. Here we consider a capacitor structure shown in Figure 4a. A monolayer of MoTe$_2$ is deposited on top of a dielectric layer of thickness *d*, which we take to be HfO$_2$ with a large dielectric constant of 25[33]. Monolayer and dielectric are sandwiched between two metal plates between which a voltage *V* is applied. High dielectric constant material HfO$_2$ is chosen to increase the capacitance and hence increase the charge density in the monolayer. The metal plate is chosen to be Aluminum with a work function of 4.08 eV. The curves in Figure 4 assume the monolayer to be at a state of constant stress, with both 2H and 1T' phases structurally relaxed. We compute the total energy and equilibrium grand potential of this system using equations (1) – (5).

Plotted in Figure 4b is the total energy (equation 1) of the capacitor shown in Figure 4a as a function of charge density in monolayer MoTe$_2$. Two black dashed lines depict common tangents between 2H and 1T' energy surfaces, the slopes of which are defined by the set of equations,

$$\left(\frac{\partial E_H}{\partial Q}\right)_{Q_H} = \left(\frac{\partial E_{T'}}{\partial Q}\right)_{Q_{T'}} = \frac{E_H(Q_H) - E_{T'}(Q_{T'})}{Q_H - Q_{T'}} \qquad (7)$$

where $V_t = \left(\frac{\partial E_H}{\partial Q}\right)_{Q_H} = \left(\frac{\partial E_{T'}}{\partial Q}\right)_{Q_{T'}}$ is the common transition gate voltage.

Plotted in Figure 4c is the equilibrium grand potential (equation 5) as a function of the gate voltage. Two transition voltages are labeled also using black dashed lines. Figures 4b and 4c show that a transition gate voltage of -1.8 V or 4.4 V can be applied to drive the phase transition in monolayer MoTe$_2$ using the capacitor in Figure 4a. The experimental breakdown voltage for a 4.5 nm-thick HfO$_2$ is reported to be as large as 3.825 V[34], which is larger than twice the magnitude of the negative transition voltage. This breakdown field in HfO$_2$ is larger than some other reports and may depend on the details of growth[35,36]. Therefore, employing an appropriate dielectric is likely to be critical here in observing the phase change. Ionic liquids may be employed to help address the challenge of achieving large electric fields.

While the curves in Figure 4 assume the monolayer is at a state of zero stress across the transition, Figure 5 presents calculations for MoTe$_2$ at constant stress (Figure 5a) and constant area (Figure 5b) utilizing the capacitor structure shown in Figure 4a. These phase diagrams predict the thermodynamically favored phase as a function of voltage $V$ and thickness $d$ of the HfO$_2$ dielectric medium. In each phase diagram, there exist two phase boundaries, the positions of which vary with the work function $W$ of the capacitor plate. The 2H semiconducting phase of MoTe$_2$ is stable between the two phase boundaries, and metallic 1T'-MoTe$_2$ is stabilized by application of sufficiently positive or negative gate voltages. The transition voltages increase with the thickness of the dielectric layer. For a capacitor containing a HfO$_2$ dielectric layer of thickness smaller than 5 nm, a negative gate voltage of approximately -2 V may be applied to drive the semiconductor-to-semimetal phase transition at constant stress (Figure 5a) but the required voltage increases to approximately -4 V at constant area in Figure 5b. In analog with the changes in charge density phase boundaries shown in Figure 3, the voltage magnitudes for the transition are larger at constant area conditions (Figure 5b) than at constant stress (Figure 5a). If the substrate constrains the area of the monolayer across the transition through friction, the voltages in Figure 5b is expected to be applicable. The figure also shows a reported experimental breakdown voltage of a 4.5 nm-thick HfO$_2$ film[34].

Field-effect transistors based on few-layered MoTe$_2$ have been reported in Ref[37] using a 270 nm thick SiO$_2$ gate dielectric layer (3.9 dielectric constant) with gate voltages as large as -50 V. For monolayer MoTe$_2$ (rather than few layers), our model predicts that a gate voltage larger than 200 V is required to drive the phase transition for this device configuration. Both the increase of dielectric thickness (from 5 nm to 270 nm) and the decrease of dielectric constant (from 25 for HfO$_2$ to 3.9 for SiO$_2$[34]) will result in larger transition gate voltages than shown in Figure 5. To observe the 2H-1T' phase transition in a device, choosing a dielectric medium of large dielectric constant and dielectric performance will be critical.

**Reducing transition gate voltages with the alloy Mo$_x$W$_{1-x}$Te$_2$**

Monolayer alloys present the possibility for reducing the required gate voltage by varying the chemical composition. Recently, monolayer alloys of Mo- and W-dichalcogenides have attracted increasing attention for their tunable properties[38–41]. We hypothesize that the 2H-1T' transition gate voltage can be tuned to lower values by alloying MoTe$_2$-WTe$_2$ monolayers. This is because in monolayer MoTe$_2$, the 2H phase is energetically favorable by 31 m*e*V per formula unit relative to the 1T' phase, whereas in monolayer WTe$_2$, the energy of the 1T' phase is 123 m*e*V per formula lower than the 2H phase, as shown in Supplementary Figure 5. Therefore, one might expect the energy difference between the two phases to be tunable through zero with alloy composition, enabling tuning of the gate voltage.

To study this possibility, we compute the phase diagram for the monolayer alloy $Mo_{0.67}W_{0.33}Te_2$ at constant area. The 2H phase of this alloy is a semiconductor with a semilocal DFT band gap of approximately 0.9 eV, and is 15 meV higher than the 1T' phase, which is metallic or semimetallic. Figure 6 shows that the 2H-1T' phase transition in this alloy can be driven by negative gating of a smaller gate voltage than pure $MoTe_2$ monolayer. For example, assuming $HfO_2$ medium of 4.5 nm thickness and capacitor plate of 4.0 eV work function, the magnitude of negative transition gate voltage can be reduced from 3.6 V ($MoTe_2$, constant-area case) to 0.6 V in the constant-area case of $Mo_{0.67}W_{0.33}Te_2$ monolayer.

One might expect that the transition gate voltage in monolayers can be tuned and reduced potentially arbitrarily by controlling the chemical composition of this and other alloys. Alternative mechanical constraints placed on the alloy monolayer (e.g. constant stress) can also be expected to shift the phase boundary and transition gate voltage. See Supplementary Figure 6 for the computational cells we used for this alloy.

**Comparison with STM experiment results of $TaSe_2$**

While electrically induced phase changes in monolayers as described here have not been experimentally observed, we seek to compute the phase boundary for an electrically-induced structural transition on the surface of a bulk Ta-based dichalcogenide and compare with available STM experimental reports[17,18]. Zhang et al.[17] reported a solid-solid phase transition in the top layer of bulk $TaSe_2$ induced by an STM tip with a negative bias voltage applied to the tip from -1.2 V to -1.8 V. This corresponds to a gate voltage applied to the TaSe$_2$ of positive 1.2 V to 1.8 V in the schematic of Figure 7a. A similar effect was reported on the surface of TaS$_2$[18].

Figure 7b shows the computed phase diagram of monolayer TaSe$_2$ in the capacitor gating structure shown in Figure 7a. It is assumed to be at constant stress because some structural relaxation was observed after phase transition in the STM experiment[17]. Unlike MoTe$_2$, the phase diagram of TaSe$_2$ has only a phase boundary at positive gate voltage. Below the phase boundary, 2H-TaSe$_2$ is more stable; above the boundary, 1T-TaSe$_2$ has lower energy and is more stable. While the model developed here is not applicable for sufficiently small dielectric thicknesses *d*, Figure 7b shows a phase boundary at a small thickness of 0 - 2 nm obtained through linear extrapolation of the phase boundary at larger thicknesses. The red-shadowed region depicts a transition gate voltage of 1.2 - 1.8 V applied to TaSe$_2$, in which the 2H-1T phase transition was reported in the STM experiment[17]. The extrapolation of the phase boundary to small thickness is not inconsistent with the transition voltage reported in the STM experiment. Here we assume that the only effect of the STM tip is to apply a bias to the monolayer and we neglect the electrical current or heating that may occur. Purely mechanical deformations[17] or heating[42] were reported to be very unlikely to contribute to the structure transition. Because the transition gate voltage is larger than the breakdown voltage in the dielectric, a direct tunneling is expected to occur at small thickness *d*.

While all the quantitative results and STM images of the transition regions presented in Ref[17] were obtained using a negative bias voltage applied to the STM tip (that is, positive gate voltage applied to TaSe$_2$), the authors mention in a note and a figure caption that a similar transition was observed under positive bias voltage but do not provide details of the voltage or other conditions. This claim is inconsistent with the calculation of the phase diagram in Figure 7b. It is possible to speculate that there could exist additional mechanisms in the STM case such as joule heating or large electric fields that may be driving the phase transition at a positive applied voltage.

**Discussion**

The electrical dynamical control of structural phase in monolayer TMDs has exciting potential applications in ultrathin flexible 2D electronic devices. If the kinetics of the transformation are suitable, nonvolatile phase change memory[9] may be an application. One might expect 2D materials to have energy consumption advantages over bulk materials due to their small thickness. If the kinetics are sufficiently fast, another potential application may be subthreshold swing reduction in field effect transistors to overcome the scaling limit of conventional transistors[4]. Additionally, the change in the transmittance of light due to the phase transition of monolayer TMDs may be employed in infrared optical switching devices, such as infrared optical shutters and modulators for cameras, window coating, and infrared antennas with tunable resonance.

To summarize, we have identified a new mechanism, electrostatic gating, to induce a structural semiconductor-to-semimetal phase transition in monolayer TMDs. We have computed phase boundaries for monolayer MoTe$_2$, MoS$_2$, and TaSe$_2$. We discover that changing carrier density or electron chemical potential in the monolayer can induce a semiconductor-to-semimetal phase transition in monolayer TMDs. We find that a surface charge density less than -0.04 *e* or greater than 0.08 *e* per formula unit is required to observe the semiconductor-to-semimetal phase transition in monolayer MoTe$_2$ under constant stress conditions, and a significantly larger value of approximately -0.33 *e* or 0.92 *e* per formula unit is required in the monolayer MoS$_2$ case. A capacitor structure can be employed to dynamically control the semiconductor-to-semimetal phase transition in monolayer MoTe$_2$ with a gate voltage approximately 2-4 V for MoTe$_2$. These transition charges and voltages are expected to vary considerably with the nature of the mechanical constraint of the monolayer. While the electric fields required to observe the transition in MoTe$_2$ are likely near breakdown and could be challenging to realize in the lab, we find that the field amplitudes can be reduced arbitrarily by alloying Mo atoms with substitutional W atoms to create the alloy Mo$_x$W$_{1-x}$Te$_2$. To provide some insight into earlier STM experimental results, we have also calculated the phase boundary of a solid-solid phase transition in TaSe$_2$ and compared with available experimental reports.

**Methods**

**Electronic structure calculation.** DFT calculations were performed using the projector augmented-wave[43] pseudopotential implementation of the Vienna Ab Initio Simulation Package[44], version 5.3.3. In the DFT calculations, electron exchange and correlation effects are described by the GGA functional of Perdew, Burke, and Ernzerhof (PBE)[45]. Wave functions are expanded in a plane-wave basis set with a kinetic energy cutoff of 350 eV on an $18 \times 18 \times 1$ Monkhorst-Pack[46] k-point grid using a Gaussian smearing of 50 meV. All calculations were performed at 0 K temperature of the nuclei. The computational cell height along the c-axis is 36 Å. The convergence thresholds were $0.5 \times 10^{-8}$ eV per $MX_2$ formula unit and $0.5 \times 10^{-7}$ eV per $MX_2$ formula unit for electronic and ionic relaxations, respectively. Spin-orbit coupling is employed in all DFT calculations. The ionic relaxations were performed using conjugate gradient algorithm.

All calculations in this work were performed at zero ionic temperature, omitting the vibrational component of the free energy. Ref[20] has shown that inclusion of vibrational free energy and temperature would shift the phase boundaries closer to ambient conditions and lower the energy required to switch the phases. Therefore, one would expect inclusion of these effects to decrease the magnitude of the transition charge density and gate voltage calculated in this work. Also, the change of band gap width is expected to affect 2H-1T' phase boundary, as further discussed in Supplementary Information.

**References**

1. Lencer, D. *et al.* A map for phase-change materials. *Nat. Mater.* **7,** 972–977 (2008).

2. Wong, H.-S. P. *et al.* Phase Change Memory. *Proc. IEEE* **98,** 2201–2227 (2010).

3. Wong, H.-S. P. *et al.* Metal #x2013;Oxide RRAM. *Proc. IEEE* **100,** 1951–1970 (2012).

4. Zhou, Y. & Ramanathan, S. Correlated Electron Materials and Field Effect Transistors for Logic: A Review. *Crit. Rev. Solid State Mater. Sci.* **38,** 286–317 (2013).

5. Shu, M. J. *et al.* Ultrafast terahertz-induced response of GeSbTe phase-change materials. *Appl. Phys. Lett.* **104,** 251907 (2014).

6. Aetukuri, N. B. *et al.* Control of the metal-insulator transition in vanadium dioxide by modifying orbital occupancy. *Nat. Phys.* **9,** 661–666 (2013).

7. Cavalleri, A. *et al.* Femtosecond Structural Dynamics in ${\mathrm{VO}}_{2}$ during an Ultrafast Solid-Solid Phase Transition. *Phys. Rev. Lett.* **87,** 237401 (2001).

8. Kikuzuki, T. & Lippmaa, M. Characterizing a strain-driven phase transition in VO2. *Appl. Phys. Lett.* **96,** 132107 (2010).

9. Wuttig, M. & Yamada, N. Phase-change materials for rewriteable data storage. *Nat. Mater.* **6,** 824–832 (2007).

10. Lee, S.-H., Jung, Y. & Agarwal, R. Highly scalable non-volatile and ultra-low-power phase-change nanowire memory. *Nat. Nanotechnol.* **2,** 626–630 (2007).

11. Bulaevskiĭ, L. N. Structural transitions with formation of charge-density waves in layer compounds. *Sov. Phys. Uspekhi* **19,** 836–843 (1976).

12. McMillan, W. L. Theory of discommensurations and the commensurate-incommensurate charge-density-wave phase transition. *Phys. Rev. B* **14,** 1496–1502 (1976).

13. Thomson, R. E., Burk, B., Zettl, A. & Clarke, J. Scanning tunneling microscopy of the charge-density-wave structure in 1T-TaS2. *Phys. Rev. B* **49,** 16899–16916 (1994).

14. Scruby, C. B., Williams, P. M. & Parry, G. S. The role of charge density waves in structural transformations of 1T TaS2. *Philos. Mag.* **31,** 255–274 (1975).

15. Castro Neto, A. H. Charge Density Wave, Superconductivity, and Anomalous Metallic Behavior in 2D Transition Metal Dichalcogenides. *Phys. Rev. Lett.* **86,** 4382–4385 (2001).

16. Wilson, J. A. & Yoffe, A. D. The transition metal dichalcogenides discussion and interpretation of the observed optical, electrical and structural properties. *Adv. Phys.* **18,** 193–335 (1969).

17. Zhang, J., Liu, J., Huang, J. L., Kim, P. & Lieber, C. M. Creation of Nanocrystals Through a Solid-Solid Phase Transition Induced by an STM Tip. *Science* **274,** 757–760 (1996).

18. Kim, J.-J. *et al.* Observation of a phase transition from the T phase to the H phase induced by a STM tip in 1T-TaS_{2}. *Phys. Rev. B* **56,** R15573–R15576 (1997).

19. Mak, K. F., Lee, C., Hone, J., Shan, J. & Heinz, T. F. Atomically Thin MoS2: A New Direct-Gap Semiconductor. *Phys. Rev. Lett.* **105,** 136805 (2010).

20. Duerloo, K.-A. N., Li, Y. & Reed, E. J. Structural phase transitions in two-dimensional Mo- and W-dichalcogenide monolayers. *Nat. Commun.* **5,** (2014).

21. Eda, G. *et al.* Coherent Atomic and Electronic Heterostructures of Single-Layer MoS2. *ACS Nano* **6,** 7311–7317 (2012).

22. Gao, G. *et al.* Charge Mediated Semiconducting-to-Metallic Phase Transition in Molybdenum Disulfide Monolayer and Hydrogen Evolution Reaction in New 1T′ Phase. *J. Phys. Chem. C* **119,** 13124–13128 (2015).

23. Kan, M. *et al.* Structures and Phase Transition of a MoS2 Monolayer. *J. Phys. Chem. C* **118,** 1515–1522 (2014).

24. Kang, Y. *et al.* Plasmonic Hot Electron Induced Structural Phase Transition in a MoS2 Monolayer. *Adv. Mater.* **26,** 6467–6471 (2014).

25. Jiménez Sandoval, S., Yang, D., Frindt, R. F. & Irwin, J. C. Raman study and lattice dynamics of single molecular layers of MoS2. *Phys. Rev. B* **44,** 3955–3962 (1991).

26. Wang, Q. H., Kalantar-Zadeh, K., Kis, A., Coleman, J. N. & Strano, M. S. Electronics and optoelectronics of two-dimensional transition metal dichalcogenides. *Nat. Nanotechnol.* **7,** 699–712 (2012).

27. Radisavljevic, B., Radenovic, A., Brivio, J., Giacometti, V. & Kis, A. Single-layer MoS2 transistors. *Nat. Nanotechnol.* **6,** 147–150 (2011).

28. Brown, B. E. The crystal structures of WTe$_2$ and high-temperature MoTe$_2$. *Acta Crystallogr.* **20,** 268–274 (1966).

29. Keum, D. H. *et al.* Bandgap opening in few-layered monoclinic MoTe2. *Nat. Phys.* **11,** 482–486 (2015).

30. Ruppert, C., Aslan, O. B. & Heinz, T. F. Optical Properties and Band Gap of Single- and Few-Layer MoTe2 Crystals. *Nano Lett.* **14,** 6231–6236 (2014).

31. Lezama, I. G. *et al.* Indirect-to-Direct Band Gap Crossover in Few-Layer MoTe2. *Nano Lett.* **15,** 2336–2342 (2015).

32. Lin, Y.-C., Dumcenco, D. O., Huang, Y.-S. & Suenaga, K. Atomic mechanism of the semiconducting-to-metallic phase transition in single-layered MoS2. *Nat. Nanotechnol.* **9,** 391–396 (2014).

33. Ray, S. K., Mahapatra, R. & Maikap, S. High-k gate oxide for silicon heterostructure MOSFET devices. *J. Mater. Sci. Mater. Electron.* **17,** 689–710 (2006).

34. Kang, L. *et al.* Electrical characteristics of highly reliable ultrathin hafnium oxide gate dielectric. *IEEE Electron Device Lett.* **21,** 181–183 (2000).

35. Lee, J.-H. *et al.* Characteristics of ultrathin HfO2 gate dielectrics on strained-Si0.74Ge0.26 layers. *Appl. Phys. Lett.* **83,** 779–781 (2003).

36. Wolborski, M., Rooth, M., Bakowski, M. & Hallén, A. Characterization of HfO2 films deposited on 4H-SiC by atomic layer deposition. *J. Appl. Phys.* **101,** 124105 (2007).

37. Pradhan, N. R. *et al.* Field-Effect Transistors Based on Few-Layered α-MoTe2. *ACS Nano* **8,** 5911–5920 (2014).

38. Chen, Y. *et al.* Tunable Band Gap Photoluminescence from Atomically Thin Transition-Metal Dichalcogenide Alloys. *ACS Nano* **7,** 4610–4616 (2013).

39. Zhang, M. *et al.* Two-Dimensional Molybdenum Tungsten Diselenide Alloys: Photoluminescence, Raman Scattering, and Electrical Transport. *ACS Nano* **8,** 7130–7137 (2014).

40. Kutana, A., Penev, E. S. & Yakobson, B. I. Engineering electronic properties of layered transition-metal dichalcogenide compounds through alloying. *Nanoscale* **6,** 5820–5825 (2014).

41. Tongay, S. *et al.* Two-dimensional semiconductor alloys: Monolayer $Mo_{1-x}W_xSe_2$. *Appl. Phys. Lett.* **104,** 012101 (2014).

42. Kim, P., Zhang, J. & Lieber, C. M. in *Solid State Physics* (ed. Henry Ehrenreich and Frans Spaepen) **Volume 55,** 119–157 (Academic Press, 2001).

43. Blöchl, P. E. Projector augmented-wave method. *Phys. Rev. B* **50,** 17953–17979 (1994).

44. Kresse, G. & Furthmüller, J. Efficient iterative schemes for ab initio total-energy calculations using a plane-wave basis set. *Phys. Rev. B* **54,** 11169–11186 (1996).

45. Perdew, J. P., Burke, K. & Ernzerhof, M. Generalized Gradient Approximation Made Simple. *Phys. Rev. Lett.* **77,** 3865–3868 (1996).

46. Monkhorst, H. J. & Pack, J. D. Special points for Brillouin-zone integrations. *Phys. Rev. B* **13,** 5188–5192 (1976).

**Acknowledgements**

Our work was supported in part by the U. S. Army Research Laboratory, through the Army High Performance Computing Research Center, Cooperative Agreement W911NF-07-0027. This work was also partially supported by NSF grants EECS-1436626 and DMR-1455050, Army Research Office grant W911NF-15-1-0570, and a seed grant from Stanford System X Alliance.

**Author contributions**

E.J.R., Y.L. and K.-A.N.D. designed the simulations and the framework for thermodynamic analysis; Y.L. performed the simulations and subsequent numerical data analysis; K. W. performed the preliminary simulations. E.J.R. and Y.L. interpreted the data and wrote the paper.

**Author information**

Reprints and permissions information is available at www.nature.com/reprints. The authors declare no competing financial interests. Readers are welcome to comment on the online version of the paper. Correspondence and requests for materials should be addressed to E.J.R. (evanreed@stanford.edu).

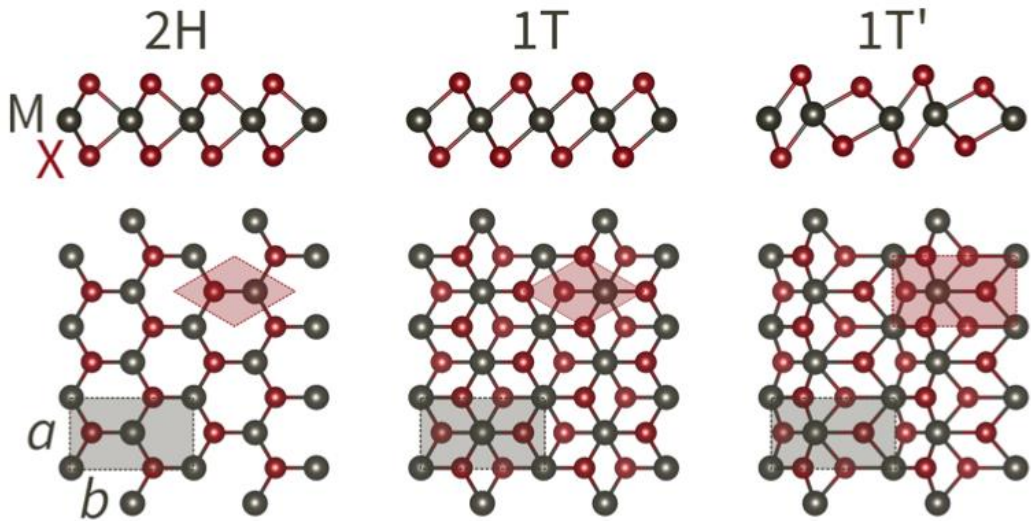

**Figure 1: Three crystal structures of monolayer TMDs.** The top schematics show cross-sectional views and the bottom schematics show basal plane views. The grey atoms are transition metal atoms and the red atoms are chalcogen atoms; in all three phases, a layer of transition metal atoms (M) is sandwiched between two chalcogenide layers (X). The semiconducting 2H phase has trigonal prismatic structure, and the metallic 1T and semimetallic 1T' phases have octahedral and distorted octahedral structures, respectively. The grey shadow represents a rectangular computational cell with dimensions $a \times b$, and the red shadow represents the primitive cell.

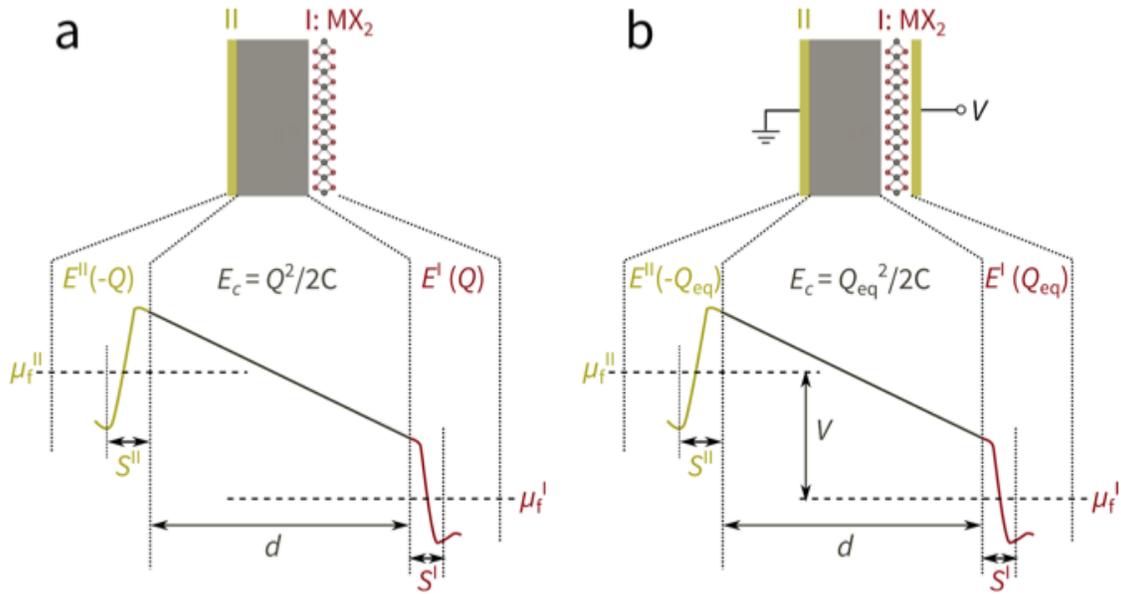

**Figure 2: Energy calculations for systems containing a charged monolayer.** Layer I is a monolayer TMD with a Fermi level $\mu_f^I$, and plate II has a Fermi level of $\mu_f^{II}$. Charge $Q$ on the monolayer TMD is fixed in panel **a**, whereas the voltage $V$ is fixed in panel **b** giving rise to an equilibrium charge $Q_{eq}$. A dielectric medium of thickness $d$ and capacitance $C$, which can be vacuum, is sandwiched between monolayer and plate. Distance $s^I$ is the separation between the center of the monolayer and the left surface of the dielectric medium, and $s^{II}$ is the separation between the surface atoms of plate II and right surface of the dielectric medium. The total energy in the fixed charge case is the sum of three parts: energy stored in the dielectric medium $E_c$, energy of moving $Q$ electrons from the Fermi level of plate II to the dielectric surface $E^{II}$, and energy of moving $Q$ electrons from the other dielectric surface to the Fermi level of monolayer TMD $E^I$.

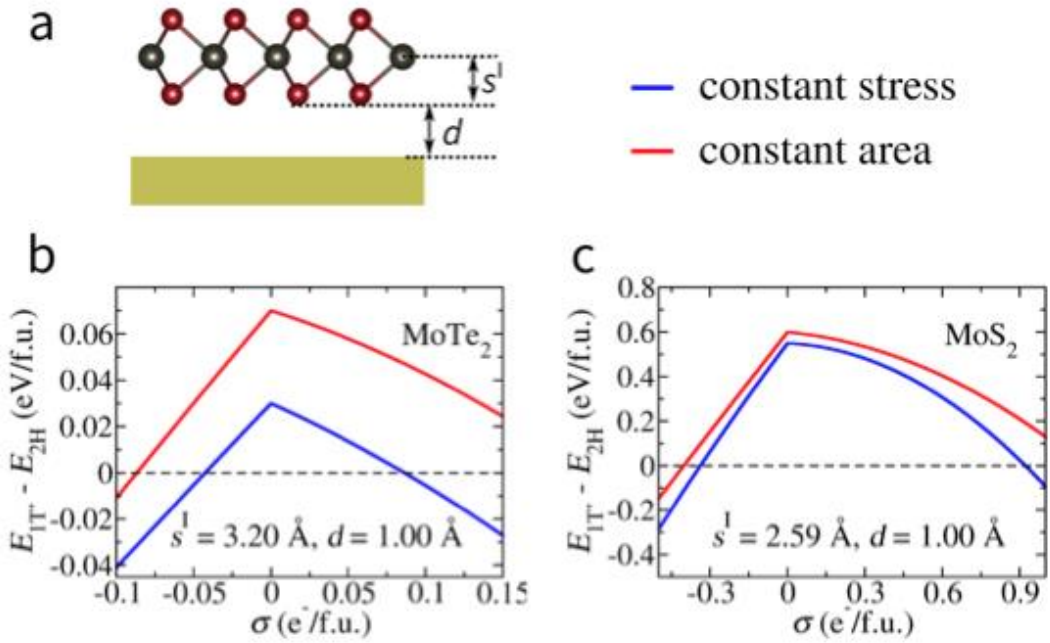

**Figure 3: Phase boundary at constant charge in monolayer MoTe$_2$ (b) and MoS$_2$ (c).** As shown in panel **a**, distance $s^l$ is the distance between the Mo atom center and the Te/S atom surface, and $d$ is the distance between the Te/S atom surface and the surface of an electron reservoir, which is chosen to be 1 Å in **b**-**c**. The internal energy difference between 2H and 1T' phases $E_{1T'} - E_{2H}$ changes with respect to the charge density $\sigma$, as shown in **b** and **c**. The blue line represents constant-stress (stress-free) case, in which both 2H and 1T' are structure relaxed. The red line represents the constant-area case, in which the monolayer is clamped to its 2H lattice constants. **b**, Semiconducting 2H-MoTe$_2$ is a stable phase and semimetallic 1T'-MoTe$_2$ is metastable when the monolayer is charge neutral or minimally charged. However, 1T'-MoTe$_2$ is more thermodynamically favorable when the monolayer is charged beyond the positive or negative threshold values. The difference in these values demonstrates a significant dependence on the relaxation of lattice constants, indicating that the precise transition point may be sensitive to the presence of a substrate. **c,** MoS$_2$ is stable in the 2H structure when charge neutral. The magnitude of charge required for the transition to 1T' is nearly an order of magnitude larger than for MoTe$_2$. In both cases, transition at constant stress is more easily induced than the transition at constant area.

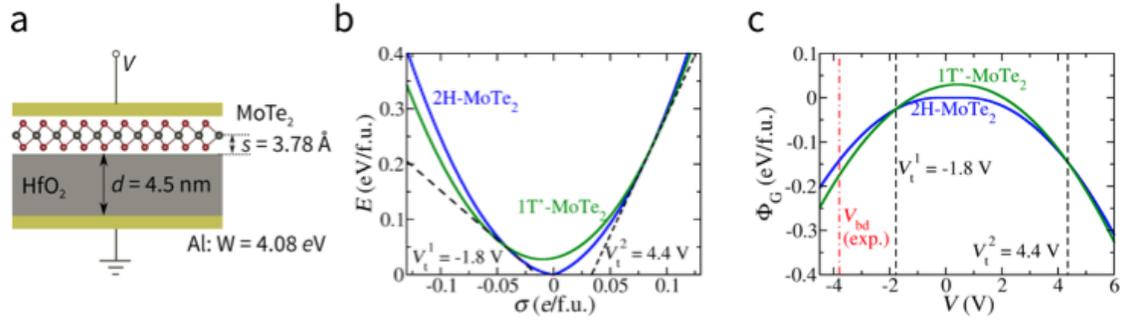

**Figure 4: Phase boundary at constant voltage.** As shown in panel **a**, monolayer MoTe$_2$ is deposited on top of a HfO$_2$ layer of thickness $d$ = 4.5 nm, and they are sandwiched between two Aluminum plates of work function $W$ = 4.08 $e$V, upon which a gate voltage $V$ is applied. The separation between MoTe$_2$ center and the surface of HfO$_2$ is assumed to be $s$ = 3.78 Å. Both 2H and 1T' are structurally relaxed (constant stress) in **b-c**. Plotted in **b** is the total energy $E$ of the capacitor shown in **a** as a function of the charge density $\sigma$ on monolayer MoTe$_2$. Plotted in **c** is the grand potential $\Phi_G$ as a function of the gate voltage $V$. The blue line represents a capacitor containing 2H-MoTe$_2$, whereas green line represents a capacitor containing 1T'-MoTe$_2$. The two black dashed lines in panel **b** represent common tangents between the 2H and 1T' energy surfaces, and in panel **c** represent intersections of the 2H and 1T' grand potentials, manifesting two transition voltages $V_t^1$ and $V_t^2$. Between the two transition voltages, semiconducting 2H-MoTe$_2$ has a lower grand potential and is thermodynamically stable. Outside this range, 1T' will be more stable. The red dashed line in panel **c** represents a breakdown voltage[34] obtained experimentally for a HfO$_2$ film of thickness 4.5 nm. The magnitude of negative transition voltage $V_t^1$ is approximately 2 V smaller than this breakdown voltage.

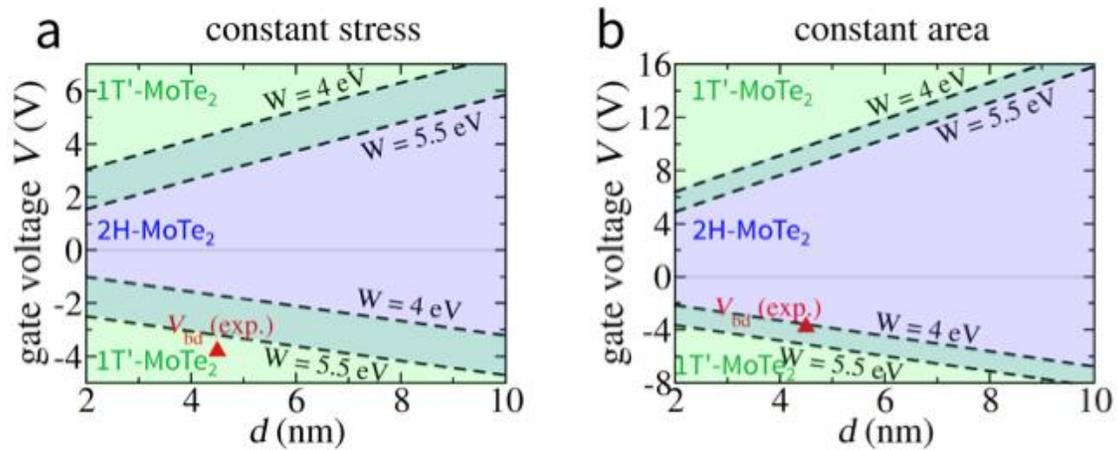

**Figure 5: Phase control of MoTe$_2$ through gating at constant stress (a) and constant area (b)**. **a-b,** Plotted are phase stabilities of monolayer MoTe$_2$ with respect to gate voltage *V* and dielectric thickness *d* using the same capacitor structure as shown in Figure 5(a). In each phase diagram, there exist two phase boundaries, which vary in position with the work function *W* of the capacitor plate. Between the two phase boundaries, semiconducting 2H is more stable, and outside 1T' is the stable structure. The phase boundaries are closer (i.e., the required transition gate voltage is smaller) in constant-stress case (**a**) than in constant-area case (**b**). For a constant-stress scenario, a negative gate voltage as small as -1 to -2 V can trigger the semiconducting-to-semimetallic phase transition in monolayer MoTe$_2$. The red triangle represents the breakdown voltage of a 4.5 nm-thick HfO$_2$ film obtained experimentally[34], which is outside the edge of the negative phase boundary.

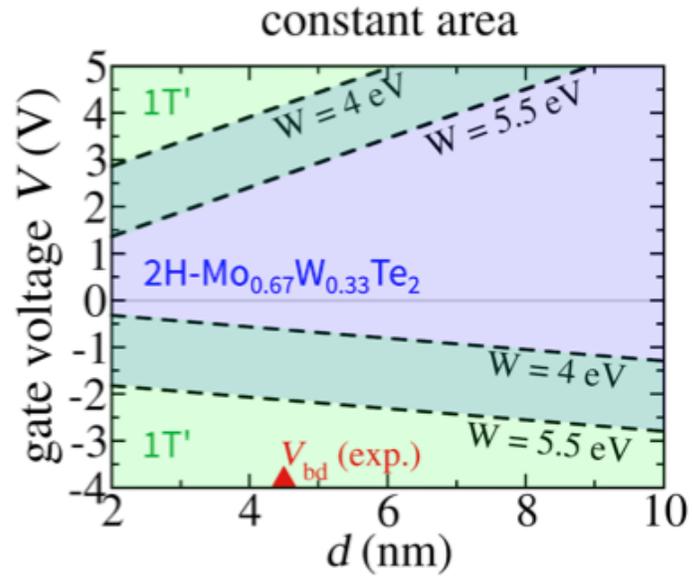

**Figure 6: Reducing transition gate voltages with the alloy Mo$_x$W$_{1-x}$Te$_2$.** Plotted is the phase stability of alloyed monolayer Mo$_{0.67}$W$_{0.33}$Te$_2$ with respect to the gate voltage $V$ and the dielectric thickness $d$ using the capacitor structure as shown in Figure 5(a). The transition is assumed to occur at constant monolayer area. The magnitudes of the transition gate voltages in this alloy are smaller than those of pure MoTe$_2$ monolayer.

## Supplementary Information

**1. Energy calculations of charged monolayers using DFT.**

In a DFT simulation of a charged monolayer with periodic boundary conditions using Vienna Ab Initio Simulation Package (VASP)[1], a homogeneous background charge is automatically introduced in the vacuum space in order to compensate for the excess charge, allowing for the periodic computational cell to remain electrically neutral and the total electrostatic energy to remain finite. A reference plane parallel to the monolayer is chosen, at a distance $z_{ref}$ away from the monolayer center, as shown in Supplementary Figure 1. The energy of moving charge $Q$ from the reference plane to the Fermi level of the monolayer $E^{mo}(Q, z_{ref})$ is calculated from

$$E^{mo}(Q, z_{ref}) = E_0 + \int_0^Q \Delta V(Q', z_{ref}) dQ' \qquad (1)$$

where $E_0$ is the ground-state energy of an electrically neutral monolayer, and $\Delta V(Q', z_{ref})$ is an appropriately defined potential difference between the reference plane $z_{ref}$ and the surface of the material defined as $z_f$. We take $z_f$ to be defined as the plane at which the plane-averaged Kohn-Sham potential is equal to Fermi level, as shown in Supplementary Figure 1. This position changes with the charge in the monolayer, i.e. $z_f = z_f(Q')$, defined via:

$$V_{tot}\left(Q', z_f(Q')\right) = \mu_f(Q') \qquad (2)$$

The potential difference $\Delta V(Q', z_{ref})$ in equation (1) is,

$$\Delta V(Q', z_{ref}) = [V_{tot}(Q', z_{ref}) - \mu_f(Q')] - [V_{bg}(Q', z_{ref}) - V_{bg}(Q', z_f(Q'))] +$$

$$\frac{Q'}{2\varepsilon_0 A}[z_{ref} - z_f(Q')] \quad (3)$$

where $V_{bg}$ is the electrostatic potential generated by the uniform compensating background charge $-Q'$ and $A$ is the area of the monolayer.

The background potential $V_{bg}$ in equation (3) is given by

$$V_{bg}(Q', z) = \frac{Q'}{2\varepsilon_0 AL}\left(z^2 - \frac{1}{4}L^2\right) \quad (4)$$

where $L$ is the size of the computational cell along $c$, interlayer axis. Equation (4) assumes that the monolayer is placed at $z = 0$.

The third and fourth terms in equation (3) are included to subtract the effect of the compensating background charge. Supplementary Figure 1 shows that after subtracting the background charge effect, a uniform electric field remains on both sides in the vacuum, which is caused by a charged monolayer, as one would expect. However, the magnitude of this electric field is only half of the electric field inside a parallel capacitor with charge $Q'$ on each plate. Therefore, the last two terms in equation (3) are added to restore the magnitude of the electric field in a parallel capacitor consisting of a monolayer charged with $Q'$.

The process of obtaining an analytical expression for the energy of a charged monolayer $E^{mo}(Q, z_{ref})$ in equation (1) can be described as follows:

First, DFT simulations are performed for a monolayer charged with different values of $Q'$. For monolayer TMDs, $Q'$ is chosen to vary from -0.06 $e$/MX$_2$ to 0.15 $e$/MX$_2$ with an increment of 0.01 $e$/MX$_2$, and a vacuum space of 36 Å is used in all cases. Secondly, a reference plane $z_0 = z_{ref}$ is chosen. For monolayer TMDs, this distance is chosen to be 17.25 Å. Thirdly, for each value of $Q'$, the potential difference between the reference plane at $z_0$ and the surface of the material $\Delta V(Q', z_0)$ can be calculated using equation (3). Fourth, first-degree polynomial fitting is performed using a linear least squares regression method to get an analytical expression for $\Delta V(Q', z_0)$ as a function of $Q'$, which can be written as:

$$\Delta V(Q', z_0) = a_1 + 2a_2'Q' \qquad (5)$$

where $a_1$ and $a_2'$ are fitting coefficients.

Fifth, the potential difference at any other reference plane $z_{ref}$, $\Delta V(Q', z_{ref})$, can be computed from equation (5) using the uniform electric field:

$$\Delta V(Q', z_{ref}) = \Delta V(Q', z_0) + \frac{Q'}{\varepsilon_0 A}(z_{ref} - z_0) = a_1 + \left(\frac{z_{ref}}{\varepsilon_0 A} + 2a_2' - \frac{z_0}{\varepsilon_0 A}\right)Q' = a_1 + \left(\frac{z_{ref}}{\varepsilon_0 A} + 2a_2\right)Q' \qquad (6)$$

where $a_2 = a_2' - \frac{z_0}{2\varepsilon_0 A}$.

Last, the energy of moving charge $Q$ from the reference plane to the Fermi level of the monolayer, $E^{mo}(Q, z_{ref})$, can be calculated using equation (1) and (6):

$$E^{mo}(Q, z_{ref}) = E_0 + a_1 Q + \left(\frac{z_{ref}}{2\varepsilon_0 A} + a_2\right) Q^2 \qquad (7)$$

In Supplementary Figure 1, we show the averaged Kohn-Sham potential along the vacuum direction $z$ of a charged monolayer, as calculated using DFT. Blue color represents the sum of ionic potential, Hartree potential, and exchange-correlation (XC) potential. Red color represents the sum of Hartree potential and ionic potential only, no XC interaction included. Solid lines show the total potential $V_{tot}$, including the potential of the charged monolayer and of the uniform compensating background charge $V_{bg}$. Blue and red dashed lines show the potential of the charged monolayer only, i.e. $V_{tot} - V_{bg}$. The black dashed line shows the Fermi level $\mu_f$ as calculated using DFT. The inclusion of XC interactions only appreciably affects the potential close to the charged monolayer and the Fermi position $z_f$. The inclusion of XC interactions does not appreciably affect the potential far away from the charged monolayer. By studying a test case as described in section 4 below, we determine that using the electrostatic part of the Kohn-Sham potential (red curves in Supplementary Figure 1) to define $z_f$ gives better agreement between direct DFT computations and predictions of the model developed here.

**2. Thermodynamic potentials in different conditions.**

All DFT calculations are performed at 0 K ionic and electronic temperature.

$$T = 0 \qquad (8)$$

In the stress-free case, both 2H and 1T' phases are structurally relaxed at a condition of constant in-plane stress, and $P = 0$. Therefore, Gibbs free energy $G$ is the relevant potential, and is written as:

$$G = U + PV - TS = U \qquad (9)$$

where $U$ is internal energy.

In the scenario of constant area, 1T' is constrained to the computational cell of the relaxed 2H phase. Therefore, Helmholtz free energy $F$ should be computed, and is written as:

$$F = U - TS = U \qquad (10)$$

Because free energy is the same as internal energy in either case, they are denoted with the same symbol $E$ in main text, although it represents different thermodynamic potentials in different conditions as discussed here.

**3. Distance parameters of TMDs.**

The distance between a monolayer TMD and the substrate upon which it sits is determined by the nature of the interaction between them. When identifying appropriate parameters for monolayer TMDs, one can refer to distance parameters in bulk TMDs. Some distance parameters in bulk TMDs in the 2H phase are labeled in Supplementary Figure 2. Parameter $c$ is the distance between the centers of two neighboring layers. Parameter $t$ is the distance from the transition metal atom centers to the chalcogenide atom centers. Parameter $s_0$ is the distance from the transition metal atom centers of one layer to the chalcogenide atom surfaces of the neighboring layer. Parameter $r$ represents the distance from the chalcogenide atom centers to the chalcogenide atom surfaces, and is estimated from empirically determined atomic radii[2]. The values used for these parameters are listed in Supplementary Table 1. The parameters for 1T'/1T phases are taken to be the same as those for 2H phase.

As shown in Figure 2 of main text, the choice of distance parameter $s^I$ is not unique in the energy calculation of a system consisting of a charged monolayer. If the dielectric medium is vacuum (as shown in Figure 3 of main text), $s^I$ is chosen to be $t + r$ (see values in Supplementary Table 1). If the dielectric medium is not vacuum (as in Figures 4-6 of main text), $s^I$ is chosen to be $s_0$ (see Supplementary Table 1).

**4. Test case of a monolayer MoTe$_2$ on top of a Cs substrate.**

In the main text, we have discussed how to calculate the energy of a system consisting of a charged monolayer, and in the first section of this Supplementary Information we have discussed energy calculations of charged monolayers in DFT with periodic boundary conditions. To test and obtain confidence intervals for the approach we have proposed, we have simulated a system consisting of a monolayer MoTe$_2$ placed on a Cs substrate. The metal substrate is chosen to be Cs because Cs has a low work function to facilitate the transfer of charge from the substrate to the monolayer MoTe$_2$.

Each computational cell includes 4 formula units of MoTe$_2$ and 6 Cs atoms, with a cell size of 60 Å in the direction perpendicular to the monolayer. Supplementary Figure 3, panel a shows a side view of the computational cell (replicated in the x and y directions). The computational cell parameters are constrained to the relaxed lattice parameters of isolated MoTe$_2$, and the Cs substrate (bcc structure) is strained to fit into this computational cell. The lattice constant in the *z* direction for the Cs substrate is tuned so that the substrates for computations with the 2H and 1T' phases have the same work function of 1.9 *e*V. In these simulations, atom coordinates are fixed and only electronic optimization is performed. The strain energy of the Cs substrate is subtracted when considering the energy difference of a system consisting of 2H-MoTe$_2$ and a system consisting of 1T'-MoTe$_2$.

Plotted in Supplementary Figure 3, panel b is the averaged electrostatic potential along the vacuum direction obtained form DFT simulation. A uniform electric field $E_{field}$ between the monolayer and the substrate is generated by the charge $Q$ transferred between them,

$$E_{field} = \frac{Q}{\varepsilon_0 A}, \qquad (11)$$

where $A$ is monolayer area.

Using the approach we have proposed to make model predictions for such a system (supplemental information section 1), we first perform separate DFT calculations for isolated monolayer MoTe$_2$ and isolated Cs substrates. The energy of a MoTe$_2$ monolayer

$E^{mo}(Q, s^I)$ and the energy of a Cs substrate $E^{sub}(-Q, s^{II})$ can be computed using equation (7):

$$E^{mo}(Q, s^I) = E_0^{mo} + a_1^{mo}Q + \left(\frac{s^I}{2\varepsilon_0 A} + a_2^{mo}\right)Q^2 \qquad (12)$$

$$E^{sub}(-Q, s^{II}) = E_0^{sub} - a_1^{sub}Q + \left(\frac{s^{II}}{2\varepsilon_0 A} + a_2^{sub}\right)Q^2 \qquad (13)$$

where $s^I$ is the distance from the center of MoTe₂ to the surface of the uniform electric field region and $s^{II}$ is the distance from the surface Cs atom centers to the other surface of the uniform electric field region, as shown in Figure 2 of main text. Therefore, the separation between the center of MoTe₂ and the center of surface Cs atoms can be written as

$$d_{Mo-Cs} = s^I + s^{II} + d \qquad (14)$$

The total energy of the system $E(Q, d_{Mo-Cs})$ can be computed using Equation (1) of the main text:

$$E(Q, d_{Mo-Cs}) = E^{mo}(Q, s^I) + E^{sub}(-Q, s^{II}) + E_c \qquad (15)$$

Using equations (12) and (13), we can further compute:

$$E(Q, d_{Mo-Cs}) = E_0^{mo} + a_1^{mo}Q + \left(\frac{s^I}{2\varepsilon_0 A} + a_2^{mo}\right)Q^2 + E_0^{sub} - a_1^{sub}Q + \left(\frac{s^{II}}{2\varepsilon_0 A} + a_2^{sub}\right)Q^2 + \frac{d}{2\varepsilon_0 A}Q^2 = E_0^{mo} + E_0^{sub} + (a_1^{mo} - a_1^{sub})Q + \left(\frac{d_{Mo-Cs}}{2\varepsilon_0 A} + a_2^{mo} + a_2^{sub}\right)Q^2 \qquad (16)$$

As discussed in main text, the equilibrium charge transferred, $Q_{eq}$, can be computed through minimization of the total energy $E(Q, d_{Mo-Cs})$ (equivalent to the grand potential since no gate voltage is applied). After which, we can further compute the total energy at equilibrium:

$$E^{eq}(d_{Mo-Cs}) = E(Q_{eq}, d_{Mo-Cs}) \qquad (17)$$

The model prediction results from equation (16) and DFT simulation results are compared in Supplementary Figure 3, panels c and d. Plotted in Supplementary Figure 3, panel c is the difference between the simulation and model prediction for the charge transferred from the substrate to 2H-MoTe$_2$ as a function of $d_{Mo-Cs}$. Plotted in Supplementary Figure 3-d is the difference between the simulation and model prediction for the energy difference of a system consisting of 2H-MoTe$_2$ and a system consisting of 1T'-MoTe$_2$. For the blue curves, the sum of electrostatic potential and XC interaction is employed as the potential when applying equation (3) for the model prediction. For the red curves, only electrostatic potential is used in the model prediction. As shown in Supplementary Figure 3, panels c and d, at small separation $d_{Mo-Cs}$ the difference between the DFT simulation and the model prediction is smaller if only the electrostatic potential is used when applying the model. Therefore, when employing the approach we have proposed to calculate the energy of a charged monolayer in this work, only electrostatic potential is used when applying equation (3).

As shown in Supplementary Figure 3, panels c and d, the difference between the DFT simulation and the model prediction decreases with the increase of the separation $d_{Mo-Cs}$. For the difference of charge transferred to be smaller than 0.01 $e$/MX$_2$ and the energy difference to be smaller than 12 m$e$V/MX$_2$, the separation $d_{Mo-Cs}$ needs to be larger than 6.5 Å. For MoTe$_2$, $s^I \approx 3.20$ Å as discussed in section 3. For the Cs substrate, the distance $s^{II}$ is approximately the atomic radius of a Cs atom[2], that is $s^{II} \approx 2.60$ Å. With this, we can compute a lower bound of the medium thickness $d$ between the monolayer and the other plate for the model to be accurate:

$$d = d_{Mo-Cs} - s^I - s^{II} \geq 0.7 \text{ Å} \qquad (18)$$

This indicates that the model described in section 1 provides a reasonable description of charge-induced transitions for monolayers separated by approximately 1 Angstrom or more from the charge-donating electrode, including typical electrostatic gating device configurations. This is due in part to the fact that this model neglects any covalent chemical bonding between the monolayer and the substrate/electrode that could occur, except for the charge transferred.

**5. The effect of band gap width**

The band gap of 2H-MoTe$_2$ calculated using the PBE XC functional is 1.00 $e$V, smaller than the quasiparticle band gap reported using a GW correction (1.77 eV)[3]. In Supplementary Figure 4, panel a (the same as Figure 3c in main text), the top of the curve for semiconducting 2H-MoTe$_2$ is observed as a plateau of width equivalent to the band gap width. To estimate the effect of band gap width on transition voltage, we increase the band gap to 1.77 $eV$ by shifting the conduction band while fixing the valence band, and the result is plotted in Supplementary Figure 4, panel b. As shown in Supplementary Figure 4, the positive transition voltage $V_t^2$ does not change, which is expected since it is mainly determined by the valence band. The negative transition voltage $V_t^1$, however, is mainly determined by the conduction band, and its magnitude is reduced from 1.78 V to 1.20 V. It is known that the PBE functional usually underestimates band gaps. This may result in an over-estimation of the magnitude of the negative transition gate voltage throughout this work.

## 6. Vacuum electronic states

When excess electrons are assigned to the computational cell, a homogeneous positive background charge is automatically introduced in the vacuum space in order to compensate the excess charge. When the number of excess electrons is increased to some value and the vacuum separation in the direction perpendicular to the monolayer surface is bigger than some corresponding threshold, the Kohn-Sham states will begin to occupy low-lying vacuum electronic states in the center of the vacuum region, at the boundary of the computational cell. Special attention has been given to avoid the formation of these vacuum electronic states by not adding too many excess electrons.

**References**

1. Kresse, G. & Furthmüller, J. Efficient iterative schemes for ab initio total-energy calculations using a plane-wave basis set. *Phys. Rev. B* **54,** 11169–11186 (1996).

2. Slater, J. C. Atomic Radii in Crystals. *J. Chem. Phys.* **41,** 3199–3204 (1964).

3. Ding, Y. *et al.* First principles study of structural, vibrational and electronic properties of graphene-like MX2 (M=Mo, Nb, W, Ta; X=S, Se, Te) monolayers. *Phys. B Condens. Matter* **406,** 2254–2260 (2011).

4. Monkhorst, H. J. & Pack, J. D. Special points for Brillouin-zone integrations. *Phys. Rev. B* **13,** 5188–5192 (1976).

5. Böker, T. *et al.* Band structure of MoS2, MoSe2, and α-MoTe2: Angle-resolved photoelectron spectroscopy and ab initio calculations. *Phys. Rev. B* **64,** 235305 (2001).

6. Bjerkelund, E. *et al.* On the Structural Properties of the Ta(1+x)Se2 Phase. *Acta Chem. Scand.* **21,** 513–526 (1967).

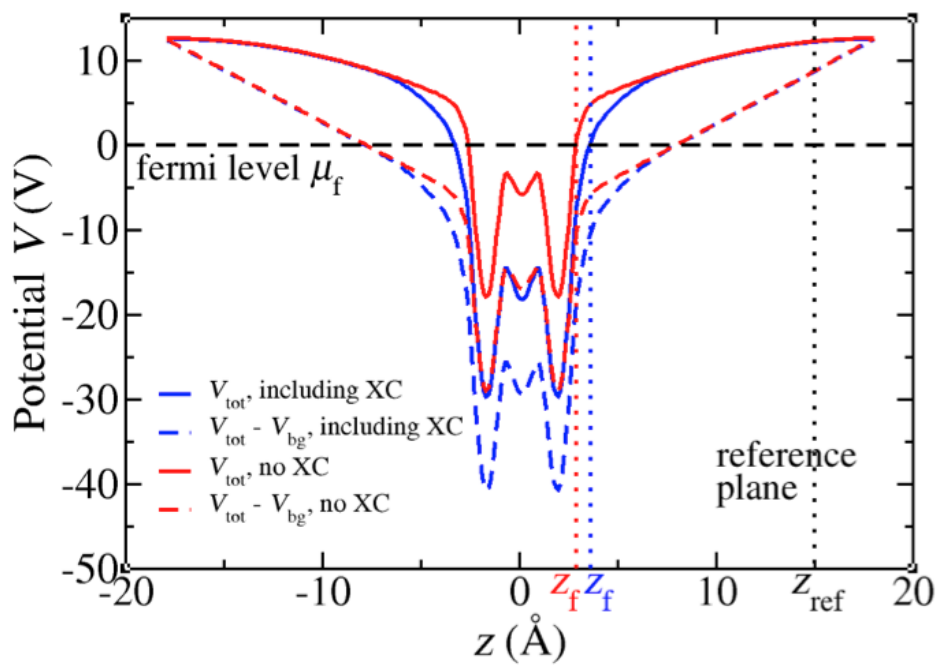

**Supplementary Figure 1: Electronic Kohn-Sham potential profile of a charged monolayer calculated using PBE-DFT.** Plotted is the averaged electronic Kohn-Sham potential along the vacuum direction $z$ of a charged monolayer, as calculated using DFT. Blue color represents the sum of electrostatic potential (ionic plus Hartree potential) and XC interaction. Red color represents the result of electrostatic potential only, no XC interaction included. Solid lines show the total potential $V_{tot}$, including the potential of the charged monolayer and of the uniform compensating background charge $V_{bg}$. Blue and red dashed line show the potential of the charged monolayer only, i.e. $V_{tot} - V_{bg}$. The black dashed line shows the Fermi level $\mu_f$ as calculated using DFT. The Fermi reference position (taken to be the intersection between the Fermi level and the total potential) is a distance $z_f$ away from the center of the monolayer. A reference plane is

positioned at a distrance $z_{ref}$ away from the monolayer center. The inclusion of XC interactions only appreciably affects the potential close to the charged monolayer and the Fermi position $z_f$. The inclusion of XC interactions does not appreciably affect the potential far away from the charged monolayer.

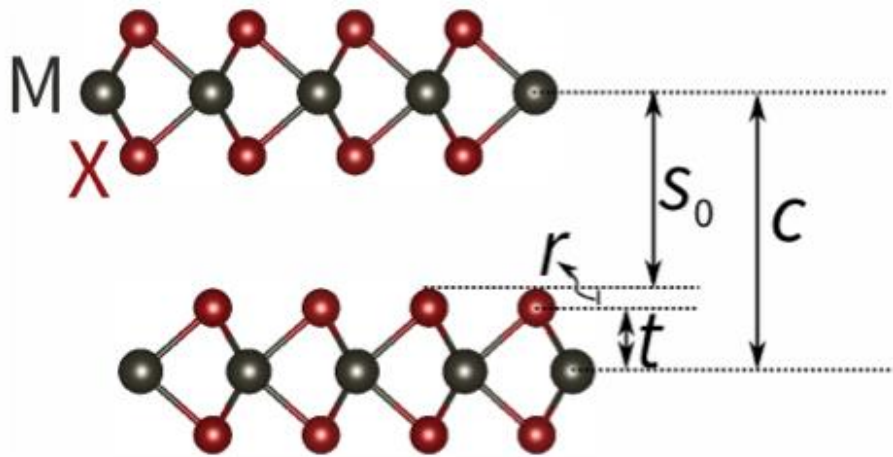

**Supplementary Figure 2: Crystal structure and distance parameters of 2H phase bulk TMDs.** Grey spheres (M) represent transition metal atoms, and red spheres (X) represent chalcogenide atoms.

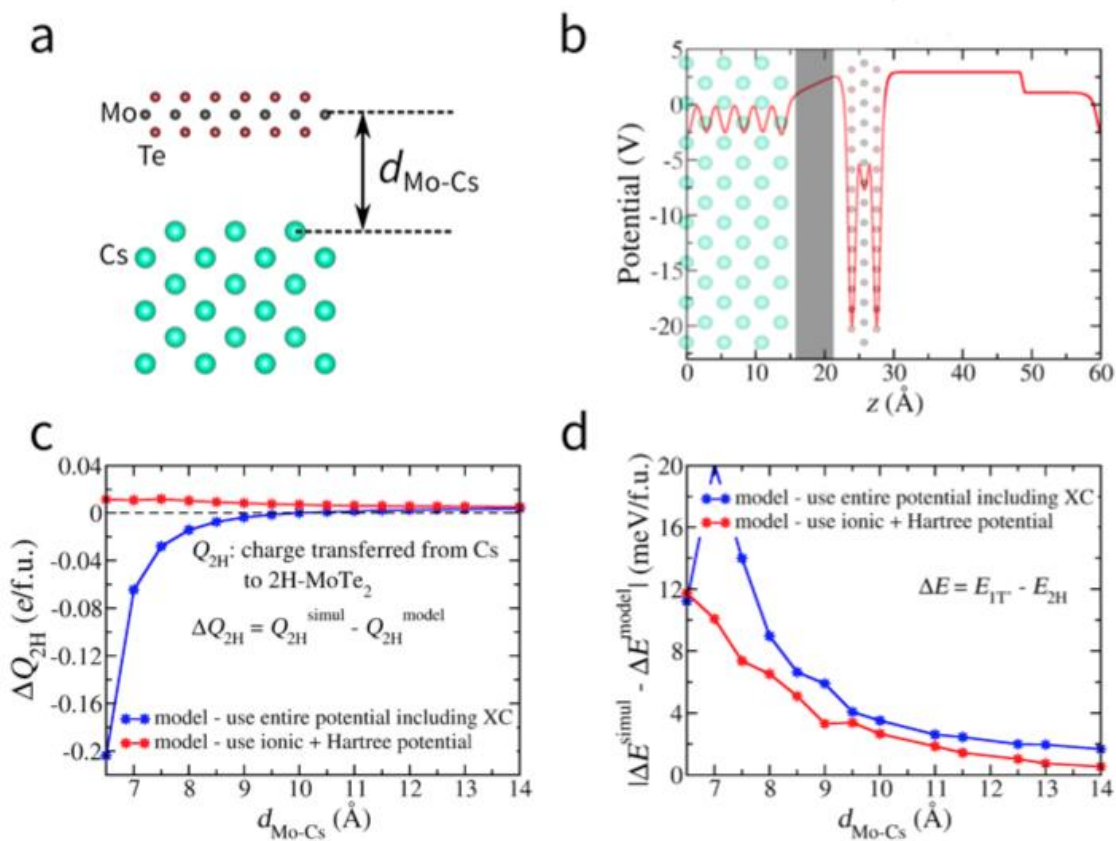

**Supplementary Figure 3: Test case of a monolayer MoTe$_2$ on top of a Cs substrate.** A system consisting of a monolayer MoTe$_2$ on top of a Cs substrate, as shown in panel **a**, is simulated using DFT and computed using the model described previously (Eqns 14-16 in Supplementary Information). The distance between the center of the monolayer and the substrate surface atom centers is $d_{Mo-Cs}$. The simulation and model prediction results are compared in panels **c** and **d**. Plotted in panel **b** is the averaged electrostatic potential of the system along the vacuum direction $z$, as calculated using DFT. Outlined in grey is a uniform electric field region, caused by the charge transferred from the Cs substrate to MoTe$_2$. Plotted in **c** is the difference between the simulation and model prediction for the charge transferred from the substrate to 2H-MoTe$_2$ as a function of $d_{\text{Mo-Cs}}$. Plotted in panel **d** is the difference between the simulation and model prediction for the energy difference of 2H and 1T' phases as a function of $d_{\text{Mo-Cs}}$. For blue curves, the potential used in the model prediction is obtained from the sum of the electrostatic potential (ionic plus Hartree potential) and XC interaction. For the red curves, the potential used in the model prediction includes only the electrostatic terms. As shown in panels **b** and **c**, for small separation $d_{\text{Mo-Cs}}$, the difference between the simulation and the model prediction is smaller if only the electrostatic potential is used in the model prediction.

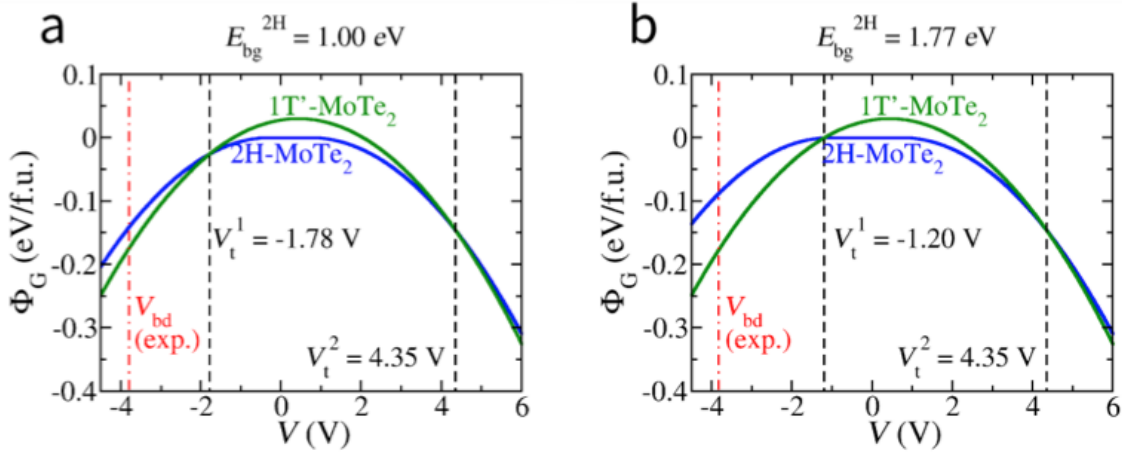

**Supplementary Figure 4: The effect of band gap on the phase boundary at constant voltage.** Panel **a** is the same as Figure 4c in main text, where the PBE-DFT band gap of 1.0 eV is used for 2H-MoTe$_2$. Panel **b** is similar to panel **a**, except that the band gap of 2H-MoTe2 is increased to 1.77 eV by shifting the conduction band. The positive transition voltage $V_t^2$ does not change, but the negative transition voltage $V_t^1$ shifts to the right and the magnitude is reduced from 1.78 V to 1.20 V.

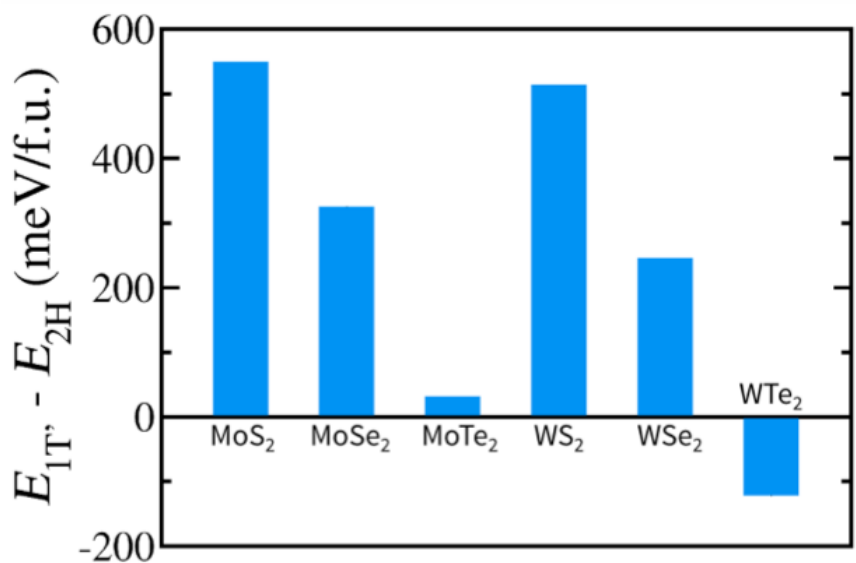

**Supplementary Figure 5: Phase energetics of 2H and 1T' monolayers**. Plotted are semilocal DFT-calculated energy differences per formula unit between freely suspended charge-neutral 2H and 1T' phases of monolayer Mo- and W-dichalcogenides, with spin-orbit coupling effects included.

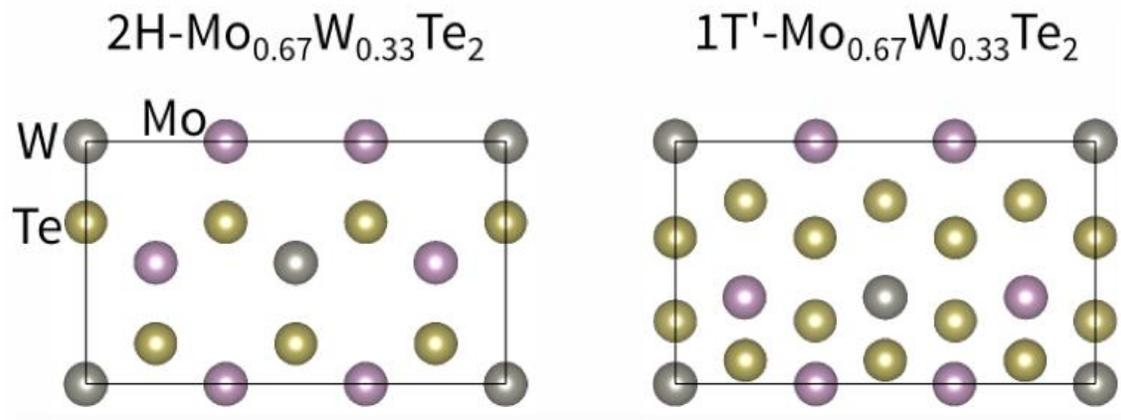

**Supplementary Figure 6: Computational cells of the crystalline phases of alloy Mo$_{0.67}$W$_{0.33}$Te$_2$ employed in this work.** Plotted are top views of computational cells of 2H and 1T' phases of alloy Mo$_{0.67}$W$_{0.33}$Te$_2$. Each computational cell consists of 12 Tellurium atoms (shown in yellow), 4 Molybdenum atoms (shown in purple), and 2 Tungsten atoms (shown in grey). The k-points are sampled on a $16 \times 16 \times 1$ Monkhorst-Pack[4] grid. A vacuum space of 36 Å is used along the z direction.

**Supplementary Table 1: Distance parameters in Å for bulk 2H-MoTe$_2$, 2H-MoS$_2$, and 2H-TaSe$_2$.** The meaning of each parameter is labeled in Supplementary Figure 2.

| Distance Parameter | 2H-MoTe$_2$ (Å) | 2H-MoS$_2$ (Å) | 2H-TaSe$_2$ (Å) |
|---|---|---|---|
| $c$ | 6.98[5] | 6.15[5] | 6.37[6] |
| $t$ | 1.80[5] | 1.59[5] | 1.67[2] |
| $r$ | 1.40[2] | 1.00[2] | 1.15[2] |
| $t + r$ | 3.20 | 2.59 | 2.82 |
| $s_0 = c - t - r$ | 3.78 | 3.56 | 3.55 |

---

[2] Calculated using PBE-DFT in this work.